\documentclass[12pt]{article} %\input epsf.tex
\usepackage{graphicx,subfigure}

\newcommand{\be}{\begin{equation}}
\newcommand{\ee}{\end{equation}}
\newcommand{\bear}{\begin{eqnarray}}
\newcommand{\eear}{\end{eqnarray}}
\newcommand{\ba}{\begin{array}}
\newcommand{\ea}{\end{array}}

\def\del{\partial}
\newcommand{\Mkk}{M_{\rm KK}}

\def\w  {\omega}

\def\ea{{\it et al}. }
\newcommand{\ie}{{\it i.e.$\,$}}
\def\ra{\rightarrow}

\newcommand{\tw}{{{\widetilde w}}}
\newcommand{\tp}{{{\widetilde p}}}
\newcommand{\tCA}{{{\widetilde C_A}}}
\newcommand{\tCB}{{{\widetilde C_B}}}
\newcommand{\tB}{{{\widetilde B}}}

\begin{document}

\begin{titlepage}
\vfill
\begin{flushright}
{\normalsize IC/2010/030}\\
{\normalsize SHEP-10-19}\\
{\normalsize arXiv:xxx.xxxx[hep-th]}\\
\end{flushright}

\vfill
\begin{center}
{\Large\bf  Holographic chiral magnetic spiral}

\vskip 0.3in

Keun-Young Kim$^1$\footnote{e-mail:
{\tt k.kim@soton.ac.uk}},
Bindusar Sahoo$^2$\footnote{e-mail:
{\tt bsahoo@ictp.it}}, and
Ho-Ung Yee$^2$\footnote{e-mail:
{\tt hyee@ictp.it}}
\vskip 0.15in

${}^1$
{\it School of Physics and Astronomy, University of
Southampton,} \\
{\it Southampton, SO17 1BJ, UK}\\[0.3in]
${}^2$
 {\it ICTP, High Energy, Cosmology and Astroparticle Physics,} \\
{\it Strada Costiera 11, 34151, Trieste, Italy}\\[0.3in]

{\normalsize  2010}

\end{center}

\vfill

\begin{abstract}

We study the ground state of baryonic/axial matter at zero temperature chiral-symmetry broken phase under a large magnetic field, in the framework of holographic QCD by Sakai-Sugimoto.
Our study is motivated by a recent proposal of chiral magnetic spiral phase that has been argued to be favored against previously studied phase of homogeneous distribution of axial/baryonic currents in terms of meson super-currents dictated by triangle anomalies in QCD.
Our results provide an existence proof of chiral magnetic spiral in strong coupling regime via holography, at least for large axial chemical potentials, whereas we don't find the phenomenon in the case of purely baryonic chemical potential.

\end{abstract}

\vfill

\end{titlepage}
\setcounter{footnote}{0}

\baselineskip 18pt \pagebreak
\renewcommand{\thepage}{\arabic{page}}
%\tableofcontents
\pagebreak

\section{Introduction and motivation }

The phase diagram of QCD with temperature and chemical potential is
an interesting area of both theoretical and experimental studies.
On the experimental side, we have an access to some part of regions through heavy-ion collision and astrophysical observations such as neutron stars. However, extracting reliable information on the state of QCD matter from these observations is not quite trivial due to theoretical difficulties
in dealing with strong interactions.
For asymptotically high temperature or chemical potential, we seem to have reliable perturbative
QCD computations to predict deconfined gas of quark/gluons and color-flavor-locked superconducting phase \cite{Alford:1998mk} respectively. However, perturbative QCD picture often seems to break down for interesting regimes such as quark-matter inside neutron stars and quark-gluon plasma in heavy-ion experiments \cite{Shuryak:2003ty}.
In fact, these situations get more complicated than the simple $(T,\mu)$ phase diagram, because they typically involve extremely large magnetic field ranging roughly $10^{12}-10^{18}$G (gauss), and
the QCD matter can feature completely new behaviors under these conditions.
Examples include magnetic catalysis \cite{Gusynin:1995nb}, anomalous axial current in dense matter \cite{Metlitski:2005pr}, chiral magnetic effect \cite{Kharzeev:2007jp}, new mechanism of pulsar kicks \cite{Charbonneau:2009ax}, spontaneous magnetization of neutron stars \cite{Gerhold:2006np}, etc to name a few.

As first-principle QCD computations are hard in most cases of interest, it is worthwhile to explore other indirect methods to attack the problems, and the AdS/CFT-inspired approach of holographic QCD is one such example. There has been recent outburst of research in this direction
applied to many aspects of QCD.
For our more specific subject of studying effects from magnetic field, previous works include Ref.\cite{Filev:2007gb,Kim,Thompson:2008qw,Bergman:2008qv,Rebhan:2008ur,Lifschytz:2009si,Yee:2009vw,Rebhan:2009vc,D'Hoker:2009bc,Gorsky:2010xu}.
One interesting feature in the studies in Refs.\cite{Thompson:2008qw,Bergman:2008qv,Rebhan:2008ur,Lifschytz:2009si,Yee:2009vw,Rebhan:2009vc,Gorsky:2010xu} is the interplay between magnetic field and triangle anomalies of global symmetries of QCD.
See also Refs.\cite{Son:2009tf,Matsuo:2009xn,Sahoo:2009yq} for hydrodynamic effects with triangle anomalies.
In all these studies, one is interested in physics of magnetic field in the deconfined phase of QCD plasma at high temperature.

The physics of QCD with magnetic field in a low temperature, chiral-symmetry broken phase\footnote{We focus on the Sakai-Sugimoto model \cite{Sakai:2004cn} where chiral-symmetry restoration happens simultaneously with deconfinement. We postpone the study of more general cases in Ref.\cite{Aharony:2006da} to the future.}
has also been of much interest in the field theory side.
One question that was addressed in Ref.\cite{Son:2007ny} is what would happen to the QCD vacuum when one introduces
small baryonic chemical potential in the presence of magnetic field.
Naively, if the baryon chemical potential is much smaller than the nucleon mass of GeV, the real
nucleon won't play a dominant role in the phase, and the physics of chemical potential might be more or less empty.
However, Ref.\cite{Son:2007ny} observed that in the presence of magnetic field, the effect from axial anomaly can change the picture. Within the low energy effective theory of chiral Lagrangian where pion field $U(x)=\exp\left( {i\over 2 f_\pi}\pi(x)\right)$ is a relevant degree of freedom, the conserved gauge-invariant baryon number current is given by
\bear B^\mu &=&
{1\over
24\pi^2}\epsilon^{\mu\nu\alpha\beta}{\rm Tr}\left(U^{-1}\partial_\nu
U U^{-1}\partial_\alpha U U^{-1}\partial_\beta U \right)\nonumber\\
&& -{1\over
24\pi^2}\epsilon^{\mu\nu\alpha\beta}\partial_\nu\left[3A^{EM}_\alpha
{\rm Tr}\left(Q(U^{-1}\partial_\beta U+\partial_\beta U U^{-1})\right)\right]\quad.\label{newbn}
\eear
where the first term is the topological Skyrmion number current representing real nucleons as Skyrmion solitons, while the second piece of our interest can be traced back to the Wess-Zumino-Witten term due to axial anomaly. It should be emphasized that the above result is uniquely determined by requiring both conservation and gauge-invariance of the current \cite{Callan:1983nx}. It is also worth of mentioning that holographic
QCD is an ideal set-up to reproduce the above result, because 5-dimensional bulk gauge field unifies
the dynamical field of pions and external gauge potential coupled to the current in a single framework
\be
A_\mu=\left[\left(U^{-1}QU\right)\psi_+ + Q\psi_-\right]A_\mu^{EM} + \psi_+U^{-1}\partial_\mu U
+({\rm excited\,\,modes})\quad,
\ee
so that the holographic version of baryon current which is {\it automatically} conserved and gauge-invariant
\be
B^\mu={1\over 8\pi^2}\int_{-\infty}^{+\infty} dz\,\epsilon^{\mu\nu\alpha\beta}\,
{\rm Tr}\left(F_{\nu\alpha}F_{\beta z}\right)\quad.\label{holocurrent}
\ee
{\it must} reproduce (\ref{newbn}), which is indeed the case \cite{Hong:2008nh}.
With the second term in (\ref{newbn}), one can have a non-zero baryon charge density with a magnetic field $B^{EM}$ and a gradient of pion $U^{-1}\partial U\sim \partial \pi$ along the same direction, {\it without} creating heavy nucleons as Skyrmions.
In the massless QCD with sufficiently small baryonic chemical potential, it has been argued that
this phase with {\it constant gradient} of pions is indeed the favored ground state.\footnote{If quarks have non-zero current masses, the pion gradient would not be constant, rather make domain walls \cite{Son:2007ny}.}
As the phase of chiral condensate $U(x)$ is rotating along the direction of pion gradient, this phase is often called {\it chiral spiral phase}. Note that this name also appears in several 1+1 dimensional models of
chiral symmetry breaking such as chiral Gross-Neveu model (2-dimensional Nambu-Jona-Lasinio model) or t'Hooft model in
the presence of chemical potential for the same reason \cite{Schon:2000he}.
In fact, it is possible to make a connection  : for sufficiently strong magnetic field, quarks are confined to the lowest Landau level and the system effectively behaves like 1+1 dimensional system.\footnote{Even without magnetic field, there have been proposals of effective 1+1 dimensional reduction on the Fermi surface of dense quark matter leading to {\it quarkyonic chiral spiral phase} \cite{Kojo:2009ha}.
We however focus on the case of low chemical potential with magnetic field in the confined phase.}

Interestingly, there recently appears a new proposal by Basar-Dunne-Kharzeev \cite{Basar:2010zd} regarding the true ground state of the system, at least for sufficiently strong magnetic field.
Based on the effective 1+1 dimensional reduction of quarks to the lowest Landau level and assuming existence of various bi-fermion condensates due to QCD interactions consistent with 1+1 dimensional Lorentz symmetry, they argue that one generically has condensates of {\it transverse currents} to the direction of magnetic field, in addition to the above mentioned chiral spirals of pion which in fact corresponds to an axial current dictated by axial anomaly.
The fermion pairings responsible for these condensates have
net non-zero momentum, so the condensate of transverse current looks like spirals and inhomogeneous.
It is important to mention that they considered both baryonic and axial chemical potentials, and anomaly dictates existence of axial/baryonic currents {\it along} the magnetic field respectively. The latter phenomenon is chiral magnetic effect \cite{Kharzeev:2007jp}.
As the transverse currents resemble chiral magnetic current {\it along} the magnetic field due to triangle anomaly, these spiral-shaped transverse currents were named as {\it chiral magnetic spirals}.

Although their arguments seem quite generic, the basic underlying assumption is the validity of quark picture which might not be clear in the confined/chiral symmetry broken phase we are considering.\footnote{In this respect, a very strong magnetic field beyond QCD scale might help because quarks can be restricted to a Landau cell whose scale can be short enough for the validity of asymptotic freedom \cite{Thompson:2008qw}. We are more interested in the regime where this is not necessarily the case, but still would like to ask the question of existence of chiral magnetic spirals.}
Another point is how to deal with strong interactions which become non-perturbative in the regime of interest.
As holographic QCD is aimed at dual description of strongly coupled regime of QCD, it seems interesting to confirm/disprove the chiral magnetic spirals within the set-up, and this is the main purpose of this paper. Our work may be considered
as a strong-coupling test of the proposal. For our objectives, it is sufficient to work in the holographic model of Sakai-Sugimoto \cite{Sakai:2004cn} as it features all the relevant chiral symmetries of QCD as well as its triangle anomalies.

In the model of holographic QCD by Sakai-Sugimoto, Ref.\cite{Thompson:2008qw,Bergman:2008qv} constructed the solution which precisely maps to the previous homogeneous chiral spiral phase. See also Ref.\cite{Rebhan:2008ur} for more study on the solution including iso-spin chemical potential as well.
Axial anomalies in QCD manifest themselves as 5-dimensional Chern-Simons terms in the holographic QCD, and indeed they play a crucial role in the existence of the solution.
An immediate generalization of their solutions is to include axial chemical potential as well as baryonic one, and this is also what we would like to consider in searching for chiral magnetic spirals. Although axial symmetry is not strictly conserved due to anomaly with QCD gluons, the violation is ${1\over N_c^2}$-suppressed in large $N_c$ limit that holographic models of QCD are based on.
A more conservative viewpoint is to take axial chemical potential or axial charge density as
a temporary, dynamical fluctuation localized in space-time. The "phase" with axial chemical potential should then be viewed as dynamical/local phase in the situation where these fluctuations are sufficiently slowly varying with respect to $\Lambda_{QCD}$.

In this more general {\it homogeneous} chiral spiral solution of the Sakai-Sugimoto model with baryonic/axial chemical potentials and the magnetic field in the chiral-symmetry broken phase, one can easily check the triangle-anomaly dictated induced currents along the magnetic field.
If the proposal of chiral magnetic spiral is valid at least in some region of parameter space of chemical potentials and magnetic field, this homogeneous chiral spiral solution in fact won't be a stable ground state against instability towards forming chiral magnetic spirals, i.e. transverse currents with net momentum. Our idea is therefore to search for any linearized instability of transverse currents in the fluctuations from the given homogeneous chiral spiral solution.
We will first give a general argument for the existence of this instability towards chiral magnetic spirals with sufficiently large axial chemical potential, and
provide numerical evidence of our phase diagram of chiral magnetic spirals in the space of baryonic/axial chemical potentials. This is at least an existence proof of the phenomenon in strong-coupling. The Chern-Simons term is essential to have such an instability, so the phenomenon is one dynamical effect of triangle anomaly. We also point out that the case of purely baryonic chemical potential which was the subject of previous studies is safely stable against the instability, so
our results indicate a closer tie between axial chemical potential and chiral magnetic spirals.
This deviates slightly from the original argument of Ref.\cite{Basar:2010zd} where both baryonic and axial chemical potentials are causing the phenomenon, and it seems interesting to understand the difference better.

It is worth of mentioning that several similar instabilities of transverse currents have previously been found in holographic models \cite{Domokos:2007kt,Nakamura:2009tf,Chuang:2010ku,Lu:2010au} and the basic mechanism is universal : strong electric field plus Chern-Simons term can induce instability \cite{Nakamura:2009tf}. In terms of mechanism, our example also belongs to the same category. However, our situation is much more delicate because of co-existence of magnetic field and electric field that are correlated with each other by equations of motion of the model. In fact, this is why it is physically more interesting due to astrophysical relevance in magnetars or spinning neutron stars and off-center heavy-ion collisions. Another realization of this aspect is the non-trivial phase diagram of chiral magnetic spirals we present in this work.

\section{Review on magnetized axial/baryonic matters in the Sakai-Sugimoto model}

In this section, we review the holographic solutions of magnetized axial/baryonic matters in the
Sakai-Sugimoto model in Refs.\cite{Thompson:2008qw,Bergman:2008qv}, which will serve as a starting point for our search for chiral magnetic spirals of transverse currents to the applied magnetic field. To be more precise, we follow Ref.\cite{Thompson:2008qw}
which constructed the solutions of two derivative truncation of the full DBI action. More elaborate
solutions of Ref.\cite{Bergman:2008qv} with the full DBI may be used in a future study.
Readers who are familiar to these solutions may skip details of this section.

The Sakai-Sugimoto model of holographic QCD is based on pairs of probe D8-$\overline{\rm D8}$ branes
for each quark flavor of QCD. These probe branes, meaning that their back-reactions to the background geometry are neglected (which is parallel to quenched approximation in lattice QCD), are embedded in some background geometry produced by $S^1$-compactified $N_c$ D4 branes. These D4 branes or the geometry they produce are responsible for $SU(N_c)$ gluonic dynamics of QCD. The dynamics of D8 branes corresponds to chiral dynamics of flavor quarks.
The description becomes simple in large $N_c$ and large t'Hooft coupling $\lambda=g_{YM}^2 N_c$ limit
where one can neglect various quantum and higher derivative corrections in this dual picture.
The model realizes the chiral symmetry breaking of massless QCD $SU(N_F)_L\times SU(N_F)_R \to
SU(N_F)_V$ in an intuitive way by adjoining D8-$\overline{\rm D8}$ branes each representing
$SU(N_F)_L$ and $SU(N_F)_R$ respectively, so that the residual symmetry on D8 branes at IR is simply $SU(N_F)_V$. One can also study finite temperature deconfined/chiral-symmetry-restored phase of the model. Relevant chiral symmetries including their axial anomalies are consistently packaged into 5-dimensional effective action on D8 branes world-volumes.

Because the main features of the physics we are interested in exist even in the case of a single flavor $N_F=1$,
we will study a single pair of D8 and $\overline{\rm D8}$ branes for simplicity, whose extension to larger number of flavors should be straightforward. This wouldn't necessarily mean that the case of multiple flavors would have same results as those results we obtain for a single flavor, because non-Abelian iso-spin symmetry might introduce a new feature that is absent in our analysis.
We also simplify our analysis by working with two derivative expansion of the full Dirac-Born-Infeld action as in Ref.\cite{Thompson:2008qw}. Although this would imply that the detailed analysis and results of the present work
have some limitation to not-too-big magnetic fields, we expect that general conclusions we draw from the present analysis won't change much in the full theory of Dirac-Born-Infeld action.
We leave these questions as open problems for the future.

The action of D8-$\overline{\rm D8}$ system up to two derivatives is \cite{Sakai:2005yt}\footnote{Greek indices span the Minkowski (3+1)-dimensions, while capital letters will be used for the full 5-dimensions. Our metric signature is $\eta=(-++++)$ and the numerical $\epsilon$ symbol is $\epsilon^{0123Z}=+1$.}
\bear
S&=&-\kappa \int dx^4 dZ\,\left[{1\over 2}\left(1+Z^2\right)^{-{1\over 3}} F_{\mu\nu} F^{\mu\nu}
+M_{KK}^2\left(1+Z^2\right) F_{\mu Z} F^{\mu Z} \right]\nonumber\\
&+&{N_c\over 96\pi^2}\int dx^4 dZ \,\epsilon^{MNPQR} A_M F_{NP} F_{QR}\quad,\label{action}
\eear
where $\kappa={\pi\over 4}\left(f_\pi\over M_{KK}\right)^2$.
We are interested in a zero temperature, chiral-symmetry broken phase where two D8 branes join at IR
to form a single D8 brane.
The second line is the Chern-Simons term responsible for triangle anomaly of $U(1)_L\times U(1)_R$
broken down to $U(1)_V$. Note that $U(1)_L\times U(1)_R$ also suffers from QCD instanton-induced anomaly with two external gluons, but it can be ignored in large $N_c$ limit in a first approximation.  The equations of motion are
\bear
\left(1+Z^2\right)^{-{1\over 3}} \partial_\mu F^{\mu\nu}+M_{KK}^2 \partial_Z\left[\left(1+Z^2\right) F^{Z\nu}\right]-{N_c\over 16\pi^2 \kappa} \epsilon^{\nu\alpha\beta\gamma}F_{Z\alpha} F_{\beta\gamma} &=& 0\,,\nonumber\\
M_{KK}^2\left(1+Z^2\right)\partial_\mu F^{\mu Z}+{N_c\over 64\pi^2\kappa}\epsilon^{\mu\nu\alpha\beta} F_{\mu\nu} F_{\alpha\beta} &=& 0\,,\label{basiceom}
\eear
and our subsequent analysis will be largely based on these only.

To realize magnetized axial/baryonic matters with an external magnetic field pointing to $x^3$-direction with a strength $B$, one looks for a solution with only $F_{0Z}$, $F_{3Z}$ and $F_{12}$ being turned on with an assumption of homogeneity $\partial_\mu=0$ at the field strength level. This is motivated by the corresponding field theory situation with baryonic chemical potential studied by Son-Stephanov \cite{Son:2007ny} where
the solution has a constant gradient of pion along $x^3$ ( in our simplified case of $N_F=1$, it would be $U(1)_A$ $\eta$-meson). Because in the Sakai-Sugimoto model, the gradient of pion field enters as
\be
\left(\lim_{Z\to\infty} Z^2 F_{3Z}+\lim_{Z\to-\infty} Z^2 F_{3Z}\right)\sim J_3^{\rm axial}\sim \partial_3 \pi +\left({\rm axial} \,\,{\rm vector}\,\,{\rm mesons}\right) \quad,
\ee
one naturally expects that the holographic solution would be homogeneous in terms of field strengths.
Note that the above statement is completely gauge-invariant, while if one chooses to work with $A_5=0$ gauge, then $\partial_\mu\pi$ will {\it also} appear in the boundary values of $\left(A_\mu(\infty)-A_\mu(-\infty)\right)$. However, one should note that this is true only in the specific $A_5=0$ gauge, and is not a correct gauge-invariant statement.
This clarification becomes important when one considers {\it both} axial and baryonic chemical potentials, as the presence of axial chemical potential will induce contributions from vector mesons too which can be read off from the field strengths only,
\be
\left(\lim_{Z\to\infty} Z^2 F_{3Z}-\lim_{Z\to-\infty} Z^2 F_{3Z}\right)\sim J_3^{\rm baryonic}\sim \left({\rm vector}\,\,{\rm mesons}\right) \quad.
\ee
The induced $J_3^{\rm baryonic}$ in terms of vector mesons due to axial chemical potential can be identified as chiral magnetic effect in the chiral-symmetry broken phase we are considering.

With this Ansatz, the equations of motion simplify as
\bear
\partial_Z\left[\left(1+Z^2\right) F_{3Z}\right] &=& -{N_c\over 8\pi^2 \kappa M_{KK}^2} F_{0Z}F_{12}\quad,\nonumber\\
\partial_Z\left[\left(1+Z^2\right) F_{0Z}\right] &=& -{N_c\over 8\pi^2 \kappa M_{KK}^2} F_{3Z}F_{12}\quad,\label{simeom}
\eear
while one of the Bianchi identities tells us
\be
\partial_Z F_{12}= \partial_2 F_{1Z}-\partial_1 F_{2Z} =0\quad,
\ee
so that $F_{12}$ is a simple constant along the radial direction $Z$,
\be
F_{12} \equiv B \quad({\rm constant})\quad.
\ee
Using this fact, the general solutions of (\ref{simeom}) are easy to be obtained as
\bear
F_{0Z}&=& {1\over 1+Z^2}\left(C_A \cosh\left[\tilde B \tan^{-1}(Z)\right]+C_B
\sinh\left[\tilde B \tan^{-1}(Z)\right]\right)\quad,\nonumber\\
F_{3Z}&=& -{1\over 1+Z^2}\left(C_B \cosh\left[\tilde B \tan^{-1}(Z)\right]+C_A
\sinh\left[\tilde B \tan^{-1}(Z)\right]\right)\quad,\label{background}
\eear
with $\tilde B={N_c B\over 8\pi^2 \kappa M_{KK}^2}$. The two integration constants $C_A$ and $C_B$ are free at the level of equations of motion, but ultimately they should be determined by two chemical potentials, axial $\mu_A$ and baryonic $\mu_B$ respectively. Note that previous literatures considered only baryonic chemical potential $\mu_B$ ( or equivalently $C_B$ ), although the above is a straightforward generalization of theirs. (\ref{background}) represents a holographic situation where
one has a magnetized matter with axial/baryonic chemical potentials {\it without} having real baryons excited. From a non-zero $F_{3Z}$, one also notices an induced axial/baryonic current along the magnetic field, and especially axial current is represented by a constant gradient of pion field.
Physically this phase will be favored against exciting real baryons, when baryonic chemical potential is smaller than the nucleon masses $\mu_B \ll M_N\sim 1$ GeV to create real baryons.
Recall that one is at zero temperature in a chiral-symmetry broken phase. The reason one can have baryonic/axial charges without having real baryons is in fact due to anomaly : external magnetic field plus an induced axial/baryonic current carry non-zero baryonic/axial charge density, respectively.
As this is from triangle anomaly of $U(1)_L\times U(1)_R$, its holographic realization is played by the 5D Chern-Simons term, as one can easily observe in the above solution.

It is not difficult to relate one integration constant $C_A$ with the axial chemical potential $\mu_A$. From the definition, one has
\bear
\mu_A&=&-{1\over 2}\left(\mu_L-\mu_R\right)=-{1\over 2}\left(A_0(+\infty)-A_0(-\infty)\right) =-{1\over 2}\int dZ\, F_{Z0}\nonumber\\
&=&-{1\over 2\tilde B} \int dZ \,\partial_Z\left[\left(1+Z^2\right) F_{3Z}\right]=
-{1\over 2\tilde B}\left[\lim_{Z\to\infty} \left(1+Z^2\right)F_{3Z}-
\lim_{Z\to-\infty} \left(1+Z^2\right)F_{3Z}\right]\nonumber\\
&=& \left[\sinh\left({\pi\over 2}\tilde B\right)\over \tilde B\right]C_A\quad,
\eear
where one has used the equation of motion (\ref{simeom}) between the first and second lines.
Hence, $C_A$ can be thought of as the axial chemical potential up to a constant factor.
Interestingly, it is much more subtle to identify correct relation between $C_B$ and the baryonic chemical potential $\mu_B$. Note that the solutions presented above do not give $\mu_B$ straightforwardly, contrary to the axial chemical potential ; $\mu_B$ as the value of ${1\over 2}(A_0(+\infty)+A_0(-\infty))$ can't be determined from the solution (\ref{background}) itself.
There have been two proposals how to relate $C_B$ with $\mu_B$. In Ref.\cite{Thompson:2008qw}, they proposed to minimize the free energy $(H-\mu_B N_B)$ with respect to $C_B$ given a fixed value of $\mu_B$, to have a relation between $C_B$ and $\mu_B$. Here $N_B$ is computed from the solution (\ref{background}) by
 the Chern-Simons induced baryon number
\be
 N_B={N_c\over 8\pi^2}\int d^3 x dZ\,\epsilon^{ijk}F_{ij} F_{kZ}={N_c B\over 4\pi^2} \int d^3x dZ\,
 F_{3Z}\quad,
\ee
while the Hamiltonian $H$ is computed from the Yang-Mills part of the action only.
Note that in this prescription, one needs only field strengths to compute things.
Another proposal in Refs.\cite{Bergman:2008qv,Rebhan:2008ur} is to extremize the on-shell action including the Chern-Simons term, instead of the previous $(H-\mu_B N_B)$. Notice that since Chern-Simons term requires gauge potential in addition to field strengths, one has to integrate the solution (\ref{background}) to get the gauge potential, and the $\mu_B$ as the value of ${1\over 2}(A_0(+\infty)+A_0(-\infty))$ enters the expressions in this way. Presumably the two prescriptions should have agreed, but curiously they give disagreeing results with each other. The basic reason for this ambiguity seems to be non gauge-invariance of the 5D Chern-Simons term.
A similar kind of issue with 5D Chern-Simons term has also been observed recently in computing zero-frequency chiral magnetic conductivity in a deconfined phase \cite{Rebhan:2009vc}.
See Refs.\cite{Gorsky:2010xu,Rubakov:2010qi,Gynther:2010ed,HUY} for proposals of the resolution.
To be conservative, we therefore leave $C_B$ as it is and present our results of analysis
in terms of $(C_B,C_A)$ parameter space. One can easily translate our results in terms of $(\mu_B,\mu_A)$
instead of $(C_B,C_A)$ after the issue will be settled down in the future. It is however
always true that $(C_{B},C_{A})$ are monotonically increasing functions of $(\mu_{B},\mu_{A})$,
so one can easily get a qualitative picture from our results.

\section{Search for holographic chiral magnetic spirals}

We come to our main motivation of looking for a signal of chiral magnetic spirals proposed by Basar-Dunne-Kharzeev \cite{Basar:2010zd}.
According to the proposal, the homogeneous solution we describe in the previous section
may {\it not} be the most favored state of the system, at least in some region of the parameter
space $(C_B,C_A)$.
One instead has  condensates of {\it transverse} axial/baryonic currents which possess a finite momentum
along the magnetic field, so that profiles of these currents form spiral shapes.
Their argument is based on an effective reduction of the system to (1+1)-dimensions in the presence of strong magnetic field with perturbative dispersion relations of
quarks/anti-quarks, although formation of various bi-fermion condensates is assumed to happen due to
QCD interactions. As QCD is strongly coupled in the chiral-symmetry broken regime, it would be valuable to check the proposal independently in the framework of holographic QCD, which can serve as an alternative proof of the existence of the phenomenon in a strongly-coupled regime.

To see whether chiral magnetic spirals are more favored than the homogeneous phase in the previous section, we study linearized stability of the homogeneous solution against small perturbations of transverse currents. In case one finds unstable modes of finite momentum along the magnetic field,
it is a sufficient proof of the existence of chiral magnetic spirals. It will also be very interesting to construct the end point of this instability, but we envision that this study may be possible only numerically and we leave this task as another future direction.

The homogeneous solutions (\ref{background}) preserve a residual rotation symmetry in the transverse $(x^1,x^2)$-plane, so that linearized fluctuations can be classified according to their helicity, and
different helicity modes are decoupled from each other.
The helicity 0 modes are fluctuations from $(A_0,A_3)$, and we are not interested in them in the present study. Instead we focus on the helicity $\pm 1$ transverse modes from $(A_1,A_2)$, which
will be responsible for a formation of chiral magnetic spirals.
Starting from the equations of motion (\ref{basiceom}), it is straightforward to obtain the linearized equations of motion governing fluctuations. Let us look at particular modes with definite frequency and momentum, $e^{-i\omega t + i p x^3}$, with {\it real} $p$ and possibly complex $\omega$ to allow instability if it exists.
Defining helicity $\pm 1$ modes as
\be
A^{(\pm)}\equiv \delta A_1 \pm i \delta A_2\quad,
\ee
they satisfy the linearized equations
\bear
\left(1+Z^2\right)^{-{1\over 3}}\left(\omega^2-p^2\right) A^{(\pm)}+
M_{KK}^2\partial_Z\left[\left(1+Z^2\right)\partial_Z A^{(\pm)}\right]\pm {N_c\over 8\pi^2 \kappa}
\left(p F_{0Z}+\omega F_{3Z}\right) A^{(\pm)} =0\,,\nonumber\\\label{linear}
\eear
where $F_{0Z}$ and $F_{3Z}$ are background solutions of (\ref{background}).
This is our master equation that we focus in the following.
Note that since we have a time-reversal symmetry $t\to -t$ in our background (this contrasts to the case of finite temperature with black-hole horizon where in-coming boundary conditions break time-reversal symmetry), a solution with $\omega$ will always accompany another solution with its conjugate $\bar\omega$.
Like a harmonic oscillator with unstable tachyonic potential, an instability can be identified
simply from having a complex-valued $\omega$ with non-zero imaginary part, so that one mode
out of $\omega$ or $\bar\omega$ grows exponentially in time.
One also observes that the last term in the equation coming from the Chern-Simons term is {\it chiral} depending on the sign of helicity, and this term will in fact be crucial to have a possible instability. The role of Chern-Simons term in causing instability was already noticed some time ago
by several authors, initiated by Domokos-Harvey \cite{Domokos:2007kt}. The basic lesson is that in the presence of sufficiently strong electric field, Chern-Simons term induces instability, which was nicely clarified
by Nakamura-Ooguri-Park recently \cite{Nakamura:2009tf}. See also Ref.\cite{Chuang:2010ku} for another realization of the instability in the Sakai-Sugimoto model in a different set-up from ours, Ref.\cite{Lu:2010au} for a generalization to higher dimensions, and Ref.\cite{Gubankova:2010ny} for 2+1 dimensional system for condensed matter application.

Perhaps, the main difference of our situation from previous cases is that we have {\it both}
electric field $F_{0Z}$ as well as magnetic fields $F_{3Z}$ and $F_{12}$, whose strengths are correlated with each other, and in fact, this correlation is precisely why the current situation is physically interesting, such as magnetars or spinning neutron stars with strong magnetic fields.
One can easily notice that what matters for instability is a Lorenz scalar $F^2$ in 5D, not electric or magnetic fields alone, so that one can get an interesting phase diagram of instabilities in the current situation, which is what we aim at in this work.

Given $(C_B,C_A)$ which enters (\ref{linear}) as $F_{0Z}$ and $F_{3Z}$, one studies an eigen-mode
problem of $\omega$ for all possible real momentum $p$ with normalizable boundary conditions on $A^{(\pm)}$ at the UV boundaries $Z\to\pm \infty$,
\be
A^{(\pm)} \sim \frac{1}{Z} \quad,\quad Z\to\pm\infty\quad. \label{Asymptotic}
\ee
The chiral magnetic spiral will be present if $\omega$ is complex for some finite range of values of momentum $p$. In general, one needs numerical analysis to map out the region of $(C_B,C_A)$ having
chiral magnetic spirals.

\begin{figure}[t]
	\centering
	\includegraphics[width=10cm]{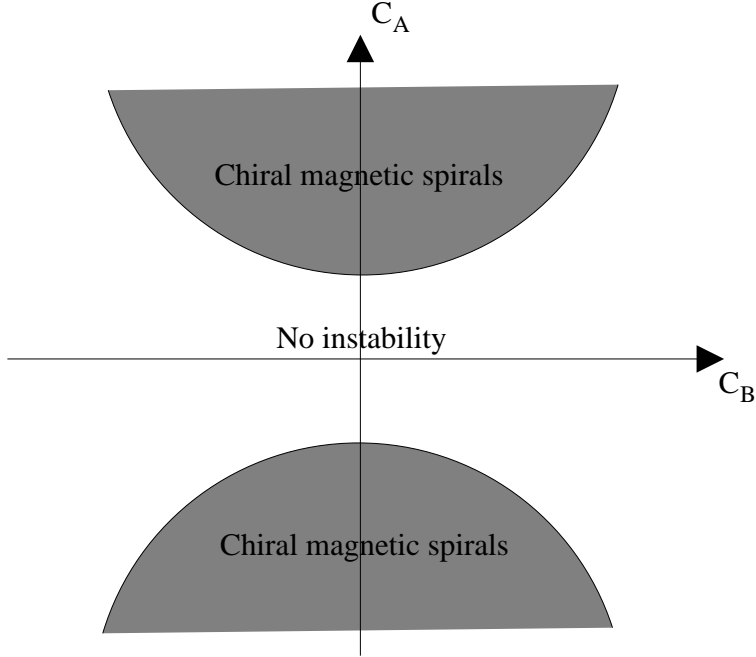}
		\caption{\label{fig1} A rough estimate of the expected phase diagram of chiral magnetic spiral in $(C_B,C_A)$ space.}
\end{figure}

Before presenting the results of our numerical study, it is possible to estimate a rough picture
how the phase diagram in $(C_B,C_A)$ would look like.
One first notices that if one removes the last term in (\ref{linear}), the equation becomes identical to the Kaluza-Klein reduction equation of massive vector mesons in the Sakai-Sugimoto model, whose
eigenvalues $\lambda^2$ of the operator
\be
\partial_Z\left[\left(1+Z^2\right)\partial_Z A\right]=-\lambda^2 \left(1+Z^2\right)^{-{1\over 3}} A\quad,
\ee
are mass squares of massive vector mesons in units of $M_{KK}^2$. From previous studies, $\lambda^2\ge \lambda_{min}^2=0.669$ where $\lambda_{min}^2$ is the mass square of the rho meson \cite{Sakai:2004cn}.
As a rough approximation, one therefore replaces this operator in (\ref{linear}) by
\be
M_{KK}^2\partial_Z\left[\left(1+Z^2\right)\partial_Z A^{(\pm)}\right]
\sim - M_{KK}^2 \lambda_{eff}^2  \left(1+Z^2\right)^{-{1\over 3}} A^{(\pm)}\quad,
\ee
with some $\lambda_{eff}^2 \ge \lambda_{min}^2$. Next, because of the UV boundary conditions $A^{(\pm)}\to 0$ as $Z\to\pm\infty$, one can assume that the wavefunctions are localized near $Z=0$ at least for low-lying states, and as a crude approximation one therefore substitutes $F_{0Z}$ and $F_{3Z}$ in (\ref{linear}) by their values at $Z=0$,
\be
F_{0Z}\sim C_A\quad,\quad F_{3Z}\sim -C_B\quad.
\ee
Then the equation (\ref{linear}) becomes an algebraic equation for $\omega$ and $p$ in this rough approximation,
\be
\left(\omega^2-p^2\right) -\lambda^2_{eff}M_{KK}^2
\pm {N_c\over 8\pi^2 \kappa}
\left(C_A p-C_B\omega \right)  =0\quad,
\ee
which can be brought into the form
\be
\left(\omega\mp {N_c C_B\over 16\pi^2\kappa}\right)^2 = \left(p\mp {N_c C_A\over 16\pi^2\kappa}\right)^2 -\left(N_c\over 16\pi^2\kappa\right)^2\left(C_A^2-C_B^2\right) +\lambda^2_{eff}M_{KK}^2\quad,
\ee
so that the eigenvalue of $\omega$ will be complex in a finite range of $p$ when
\be
-\left(N_c\over 16\pi^2\kappa\right)^2\left(C_A^2-C_B^2\right) +\lambda^2_{eff}M_{KK}^2 < 0\quad,
\label{phase}
\ee
This determines the region of $(C_B,C_A)$ having instability toward chiral magnetic spirals, which is a region defined by a hyperbola that is drawn in Figure \ref{fig1}.
We will shortly confirm numerically that this rough picture indeed captures the true phase diagram qualitatively.

It is important to notice that the purely baryonic chemical potential of $C_B\ne 0$ with $C_A=0$
is {\it stable} against chiral magnetic spirals, and in fact we mathematically prove its stability in the Appendix. Therefore, previous Son-Stephanov chiral spiral phase is not affected by our results.
This seems to indicate that axial chemical potential is the main cause of chiral magnetic spirals at strong coupling regime, while the weak coupling picture in Ref.\cite{Basar:2010zd} has both chemical potentials inducing chiral magnetic spirals. One possibility is that the dispersion relation of excitations that was used in Ref.\cite{Basar:2010zd} might be drastically modified at strong coupling in the case of baryonic chemical potential, while the case of axial chemical potential keeps its main features to have the phenomenon. We leave further understanding of this distinction to the future.

\begin{figure}[t]
	\centering
 \subfigure[Stable phase]
 {\includegraphics[width=5cm]{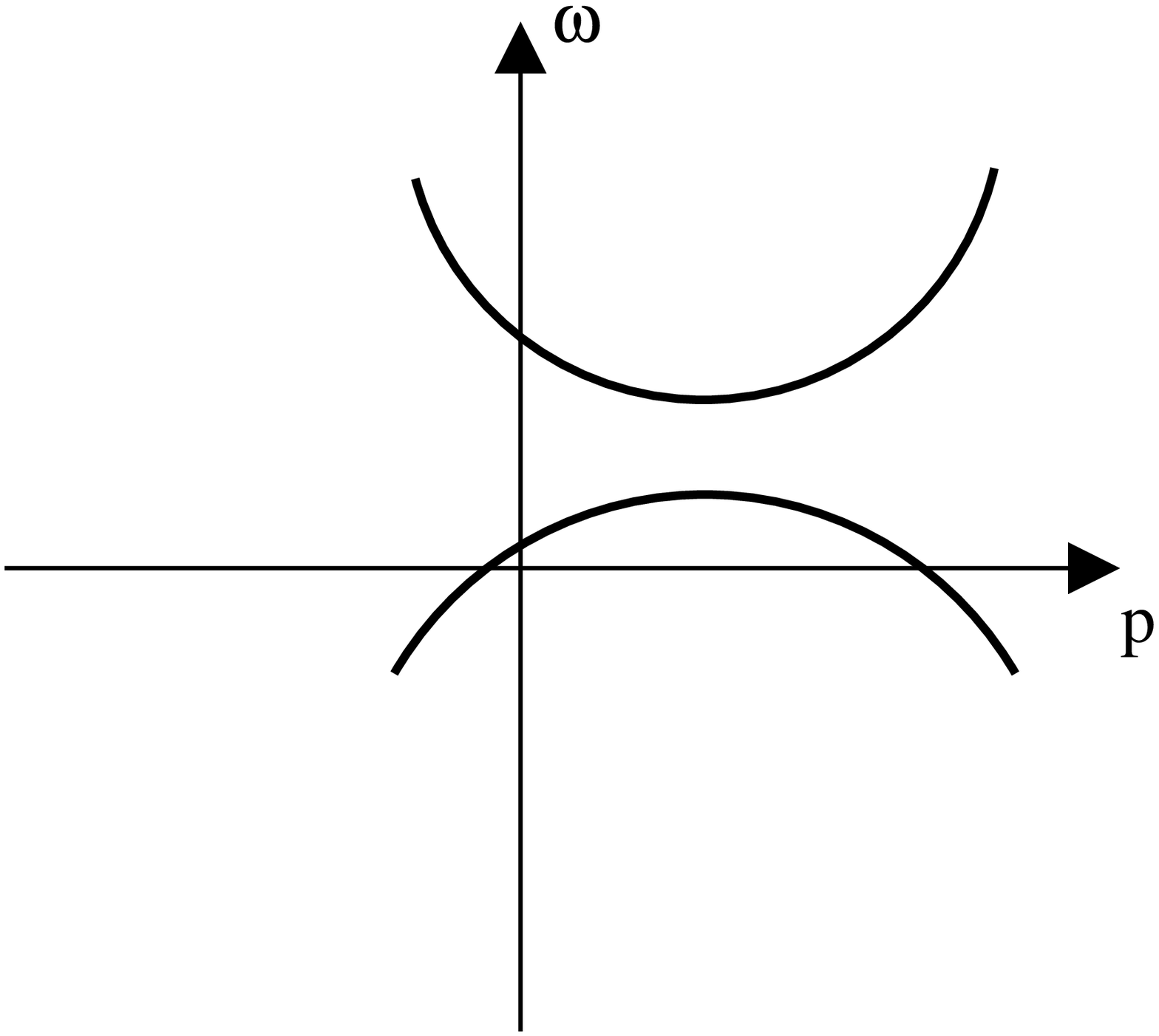}}
 \subfigure[Unstable phase]
 {\includegraphics[width=5cm]{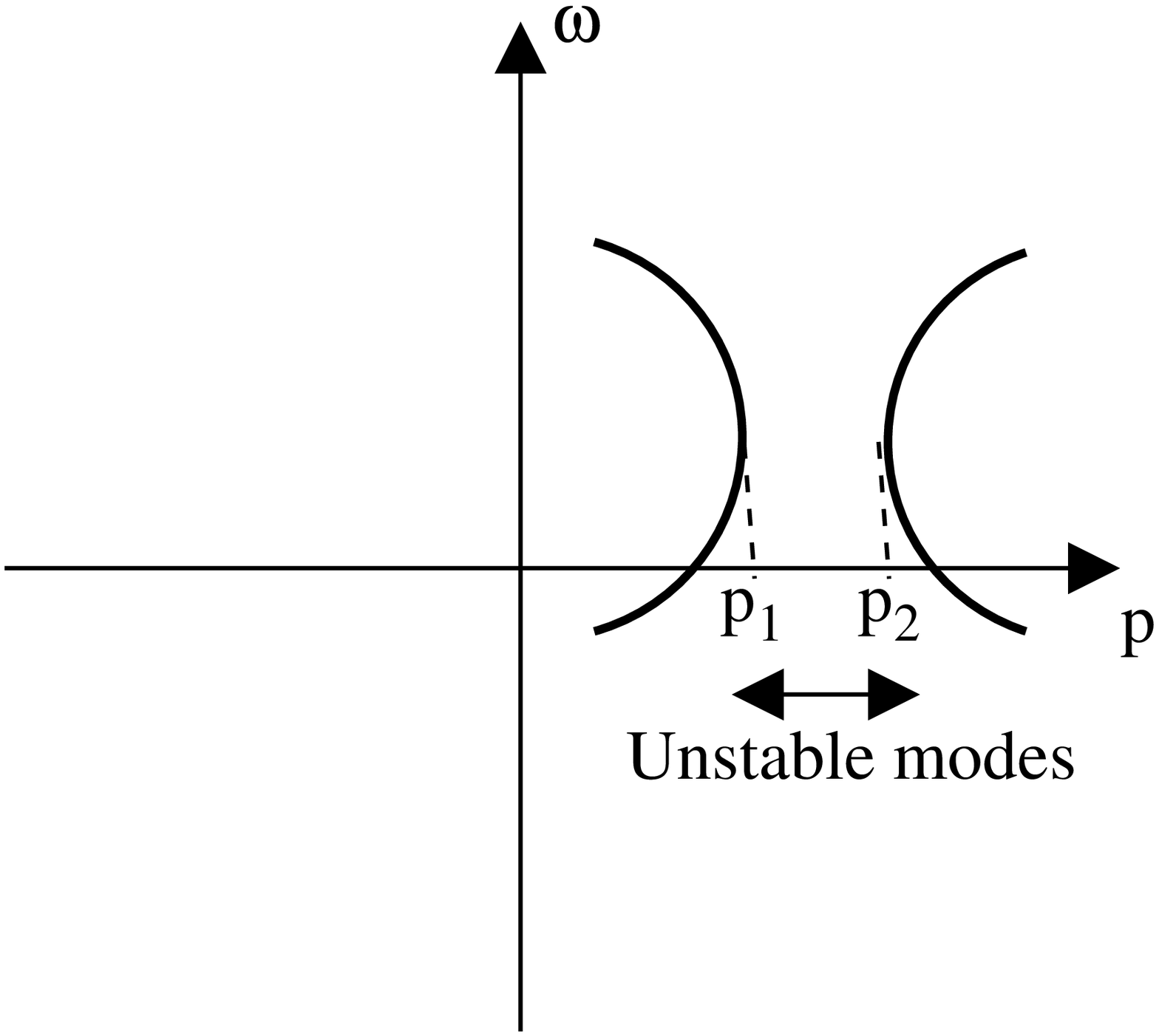}}
 \subfigure[Phase boundary]
 {\includegraphics[width=5cm]{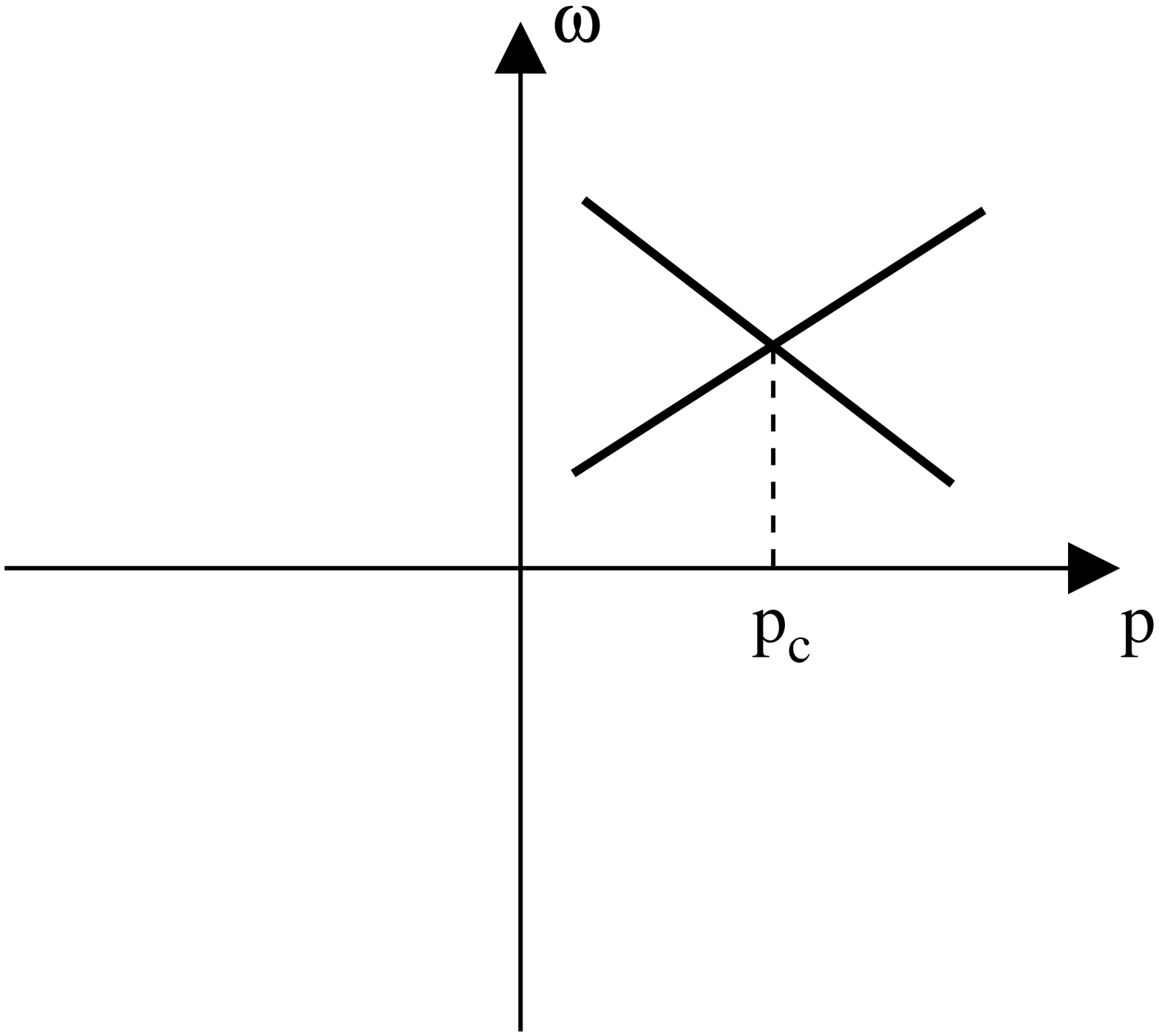}}
\caption{\label{fig2} Spectral curves in $(p,\omega)$ in the case of (a) stable phase, (b) unstable phase, and (c) at the phase boundary. }
\end{figure}
To identify the true phase boundary of our interest in our numerical study (the hyperbola in the case of the above approximation), the above rough analysis in fact provides us a very important
clue how to search for the phase boundary numerically : consider the shape of {\it dispersion relation} of {\it real} $\omega$ and $p$ as we cross the phase boundary.
In the region without chiral magnetic spirals, that is, where the condition (\ref{phase}) is not satisfied, the dispersion relation or spectral curve in {\it real} $(p,\omega)$ space looks like
a hyperbola whose transverse axis (meaning the line joining two focal points) is vertical.
In real numerical situations, the true dispersion relation will be distorted and tilted, but the
relevant feature is that the curve spans whole range of $p$ while there is a gap in $\omega$.
See Figure \ref{fig2}(a) for a schematic description. This is reasonable because we have real $\omega$ for every $p$, and there is no instability for any $p$.
In the opposite case of region with chiral magnetic spirals, that is, where (\ref{phase}) is satisfied, the dispersion relation curve would look like a hyperbola whose transverse axis is now horizontal, so that there is a gap in $p$ instead while $\omega$ is locally continuous.
For $p$ sitting between this gap, one has a complex eigenvalue for $\omega$, and hence instability to chiral magnetic spirals. See the Figure \ref{fig2}(b). At the phase boundary, one expects something like the Figure \ref{fig2}(c). What happens is that the two states with real eigenvalues of $\omega$ in Figure \ref{fig2}(a) become degenerate
at the phase boundary, and then split into two states with complex eigenvalues of $\omega$ conjugate to each other as one moves to the Figure \ref{fig2}(b). These states can no longer be seen in {\it real} $(p,\omega)$ dispersion relation we are considering.
Therefore, one can numerically distinguish the two phases by looking at dispersion relations in real $(p,\omega)$ space as one changes parameters $(C_B,C_A)$.

\begin{figure}[]
	\centering
	\subfigure[Stable phase, $\tCA = 1.9$]
	{\includegraphics[width=5cm]{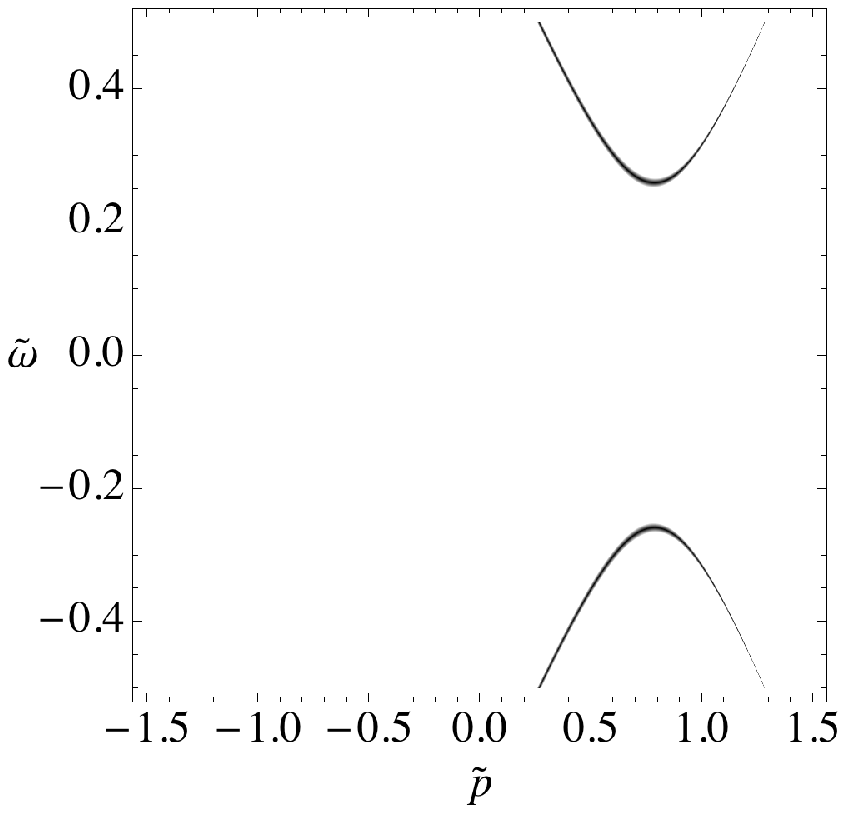}}
	\subfigure[Unstable phase, $\tCA = 2.1$]
    {\includegraphics[width=5cm]{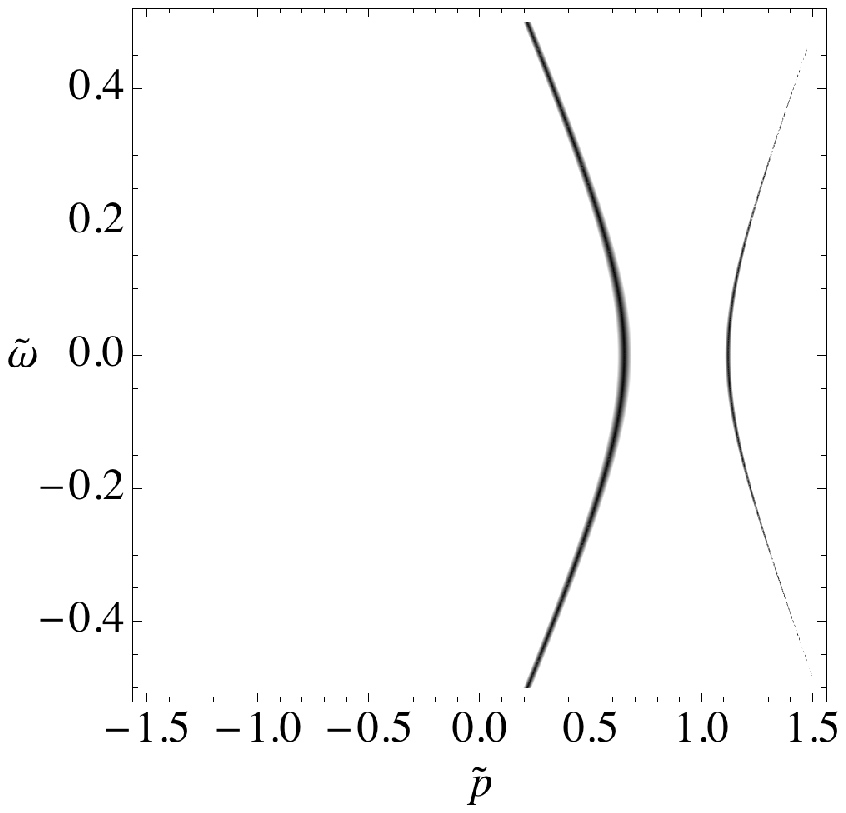}}
	\subfigure[Phase boundary, $\tCA = 2.03$]
	{\includegraphics[width=5cm]{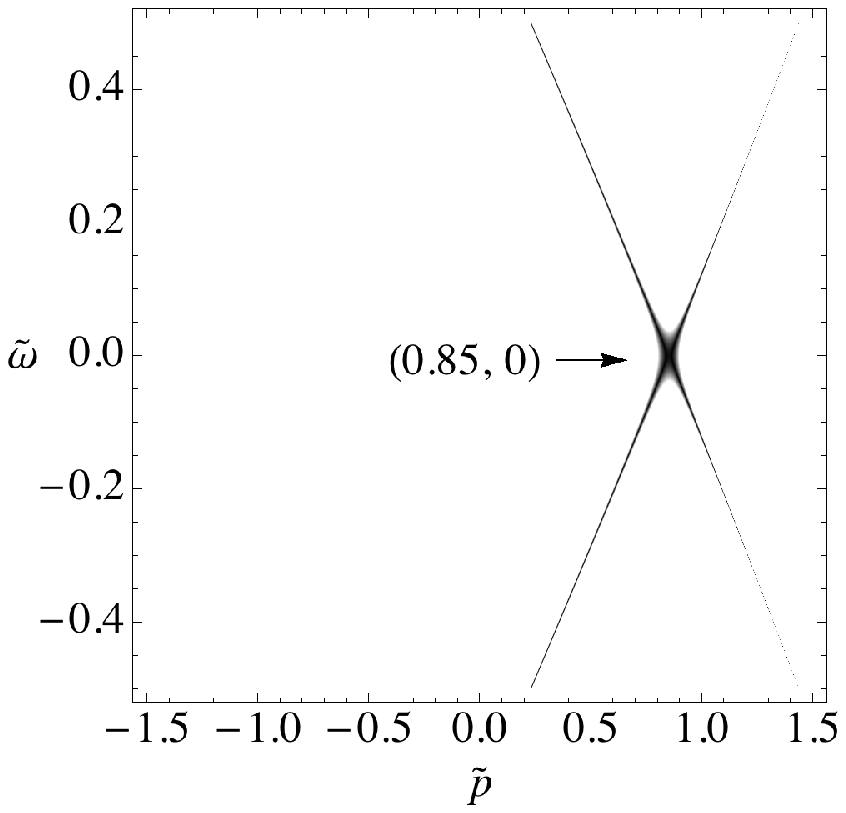}}
		\caption{\label{Fig.pw}  Spectral curves of the lightest mode in the re-scaled $(\tp,\tw)$ space with $\tB=1$ and $\tCB=0$. This confirms numerically the previous expectation in Figure \ref{fig2}.  }
\end{figure}

To illustrate how it works, we present an exemplar case of purely axial chemical potential $C_A\ne 0$ with $C_B=0$ as in Figure \ref{Fig.pw}.
%\begin{figure}[t]
%	\centering
%	\includegraphics[width=10cm]{fig5.eps}
%		\caption{\label{fig5} Three distinct points with $C_B=0$ in the phase diagram.}
%\end{figure}
For convenience, we introduce the dimensionless variables
\begin{eqnarray}
 \tp \equiv \frac{p}{\Mkk}\ , \quad  \tw \equiv \frac{\w}{\Mkk} \quad,\quad
 \left(\tCB,\tCA\right)={N_c\over 8\pi^2\kappa M_{KK}^2}\left(C_B,C_A\right)\quad,
\end{eqnarray}
so that the equation (\ref{linear}) reads as
\begin{eqnarray}
  &&(1+Z^2)^{-{1\over 3}}(\tw^2-\tp^2)A^{(\pm)} + \del_Z\left[(1+Z^2) \del_Z A^{(\pm)} \right]\pm (\tp F_{0Z} + \tw F_{3Z})A^{(\pm)} = 0\,,\nonumber\\
&&    \label{Numerical.EOM}
\end{eqnarray}
with
\begin{eqnarray}
  && F_{0Z} = \frac{1}{1+Z^2} \Big[\tCA \cosh(\tB\arctan Z) + \tCB \sinh(\tB\arctan Z)\Big] \ , \\
  && F_{3Z} = - \frac{1}{1+Z^2} \Big[\tCB \cosh(\tB\arctan Z) + \tCA \sinh(\tB\arctan Z)\Big] \ .
\end{eqnarray}
Note that $A^{(-)}$ can be obtained from $A^{(+)}$ by $\tCA \ra -\tCA$ and $\tCB \ra -\tCB$.
Also, there is a {\it parity} symmetry
\be
Z\to -Z\quad,\quad \tCB\to \tCB\quad,\quad \tCA\to -\tCA \quad,\quad \tp\to -\tp,\quad,\quad\tw\to\tw\quad,
\ee
so it is enough to study the equation of $A^{(+)}$ in the first quadrant of $\left(\tCB,\tCA\right)$
to identify the full phase diagram.
Our numerical strategy to find the spectral curves in Figure \ref{Fig.pw} is the 2D shooting method.
We shoot from the initial point at $Z_i \sim -\infty$ with the normalizable boundary condition $A^{(+)}(Z_i) = {1\over Z_i}$ in (\ref{Asymptotic}), and numerically integrate (\ref{Numerical.EOM}) to $Z_f\sim +\infty$ to find $A^{(+)}(Z_f)$. Figure \ref{Fig.pw} are simple contour plots of $\left|A^{(+)}(Z_f)\right|$ in the $(\tp,\tw)$ space where lower values are darker.
Only the right values of $(\tp,\tw)$ will lead our shooting
to be on vanishingly small values of $\left|A^{(+)}(Z_f)\right|$ at $Z_f\sim\infty$.
From the Figure~\ref{Fig.pw}(c) we can obtain one phase boundary point $(\tCB,\tCA) = (0,2.03)$.
By varying $\tCB$ we search for the corresponding $\tCA$
showing the critical behavior such as Figure~\ref{Fig.pw}(c), and this determines the phase boundary between chiral magnetic spiral phase and the stable homogeneous phase.

\begin{figure}[]
	\centering
	\includegraphics[width=10cm]{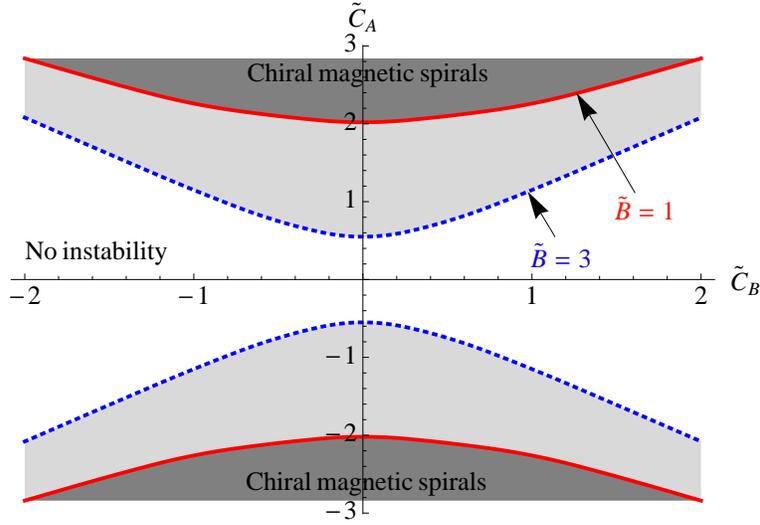}
		\caption{\label{Fig.phase} Numerical phase diagram of chiral magnetic spiral at $\tB=1$ and $\tB=3$. It conforms to our expectation that a larger magnetic field should enhance the instability  to chiral magnetic spiral.}
\end{figure}
The resulting phase diagrams of chiral magnetic spiral for two different values of magnetic field $\tB=1$ and $\tB=3$
are shown in Figure~\ref{Fig.phase}.
This is a numerical demonstration of the Figure~\ref{fig1}.
The red solid curve is for $\tB=1$ and
the blue dotted one is for $\tB=3$. The larger magnetic field enhances
the instability, so the region of chiral magnetic spirals gets bigger as expected. The full phase diagram is reflection-symmetric on $\tCA$ and $\tCB$ axis.
This is a numerical corroboration of the chiral magnetic spiral in the strong coupling regime, at least for sufficiently large axial chemical potentials.

\begin{figure}[]
	\centering
	\subfigure[Eigenvalues: numbers are $C_A$s.]
	{\includegraphics[width=6cm]{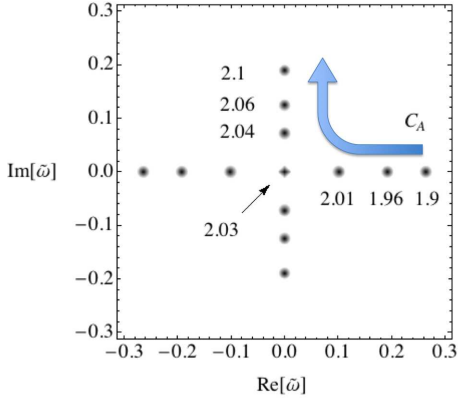}} \hspace{1cm}
	\subfigure[Eigenfunction for $C_A = 2.1, \tw=0.19i$.]
    {\includegraphics[width=7cm]{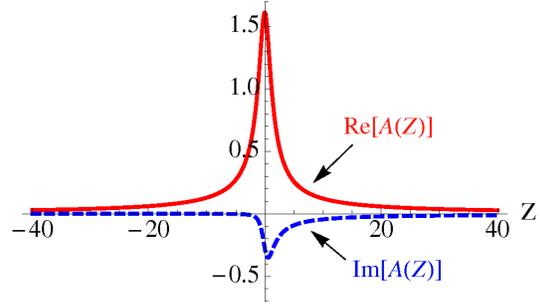}}
		\caption{\label{Fig.eigen} Eigenvalues and an explicit eigenfunction for $\tB=1, \tCB=0, \tp = 0.85$. This is a concrete numerical proof of existence of instability toward chiral magnetic spiral.}
\end{figure}

To see more explicitly how complex eigenvalues $\tw$ emerge, we trace them following the parameter space of
Figure \ref{Fig.pw}. \ie $\tB=1$, $\tCB=0$, from $\tCA = 1.9$ to $2.1$.
As an example let us choose $\tp=0.85$ which is a critical point shown in Figure \ref{Fig.pw}(c).
Since all parameters except $\tw$ are given, the eigenvalues $\tw$ can be determined
again by 2D shooting method, but now in (Re[$\tw$], Im[$\tw$]) space.
Some numerical results are plotted in Figure \ref{Fig.eigen}(a).
At $\tCA=1.9$ there are two real eigenvalues at ($\tw=\pm0.26$),
which agree to the values read from Figure \ref{Fig.pw}(a).
As $\tCA$ increases these two real eigenvalues approach to each other, and at $\tCA=2.03$ they merge at zero,
as expected from Figure \ref{Fig.pw}(c). After passing this phase boundary point ($\tCA=2.03$), new pair of
two imaginary eigenvalues appear and grow apart with increasing $\tCA$.
At $\tCA=2.1$ we find $\tw = 0.19 i$ of which complex-valued eigenfunction is shown in Figure \ref{Fig.eigen}(b).
%
%For a concrete example of eigenfunctions with complex-valued $\tw$ in an unstable regime,
%we plot the (complex) eigenfunction with $\tB=1$, $(\tCB,\tCA)=(0,2.1)$, and $\tp=0.85$ in Figure \ref{Fig.eigen}, which indeed has a complex-valued $\tw=0.19i$. This is at least an existence proof of the phenomenon in holographic QCD within the validity of our approximation.
%

%

\begin{figure}[]
	\centering
	\subfigure[$\tCA = 0$]
	{\includegraphics[width=3.5cm]{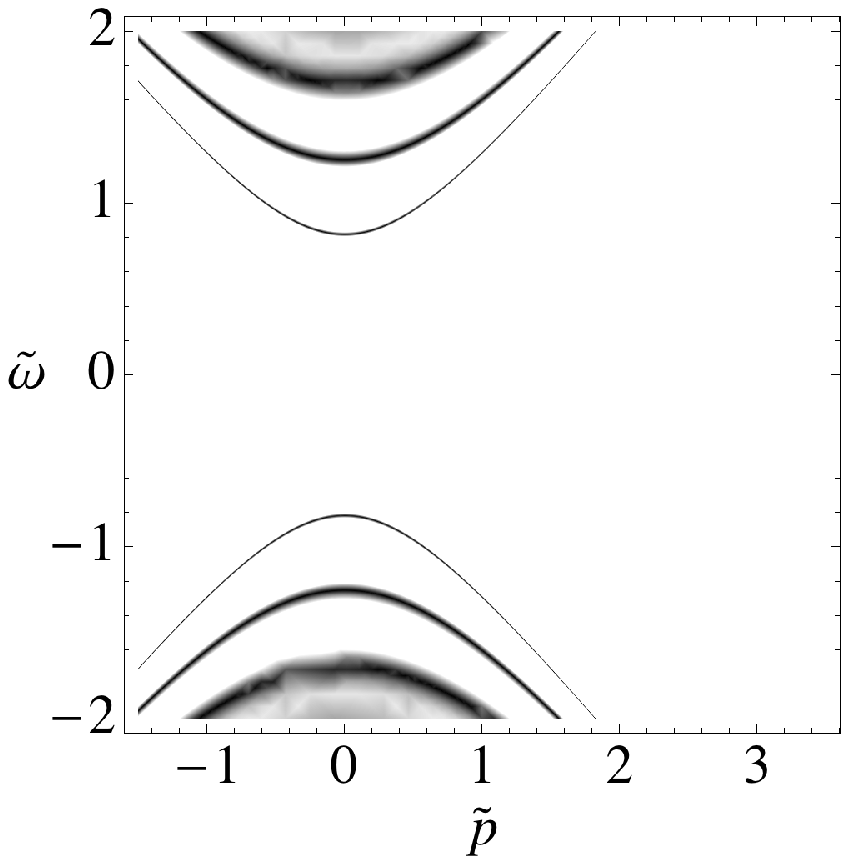}}
	\subfigure[$\tCA = 2.03$]
    {\includegraphics[width=3.5cm]{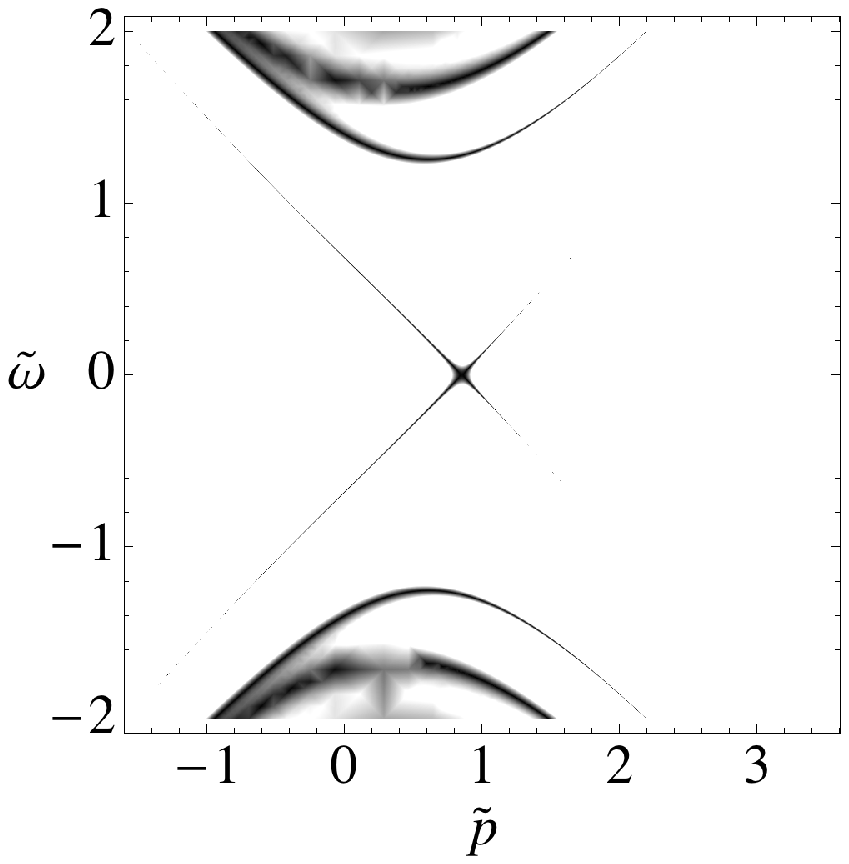}}
	\subfigure[$\tCA = 4.165$]
	{\includegraphics[width=3.5cm]{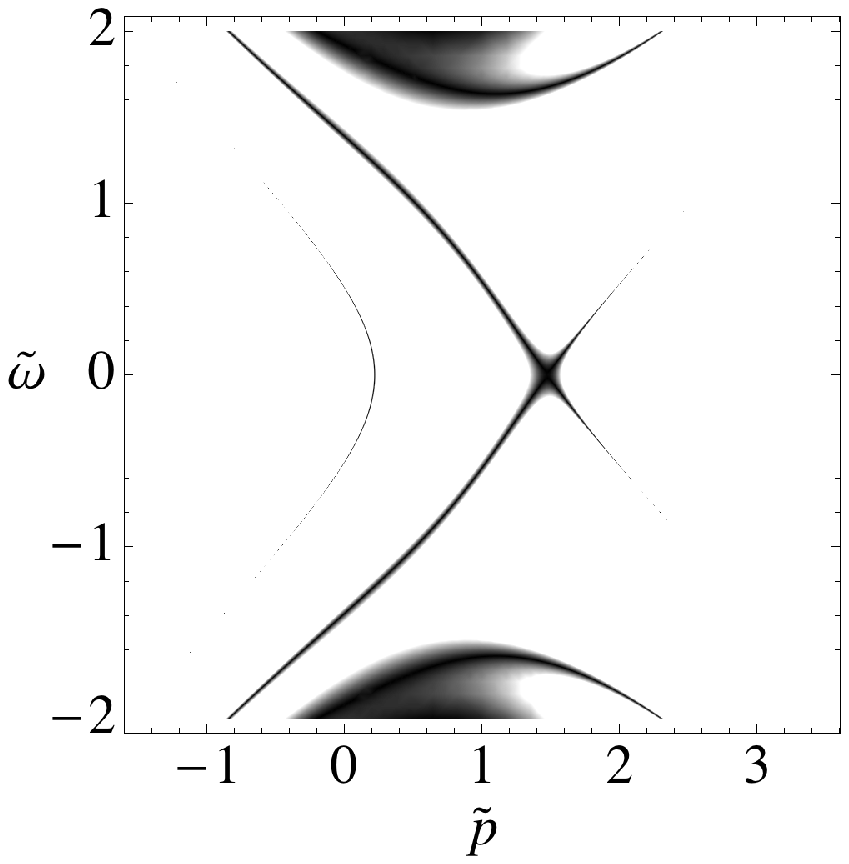}}
     \subfigure[$\tCA = 6$]
	{\includegraphics[width=3.5cm]{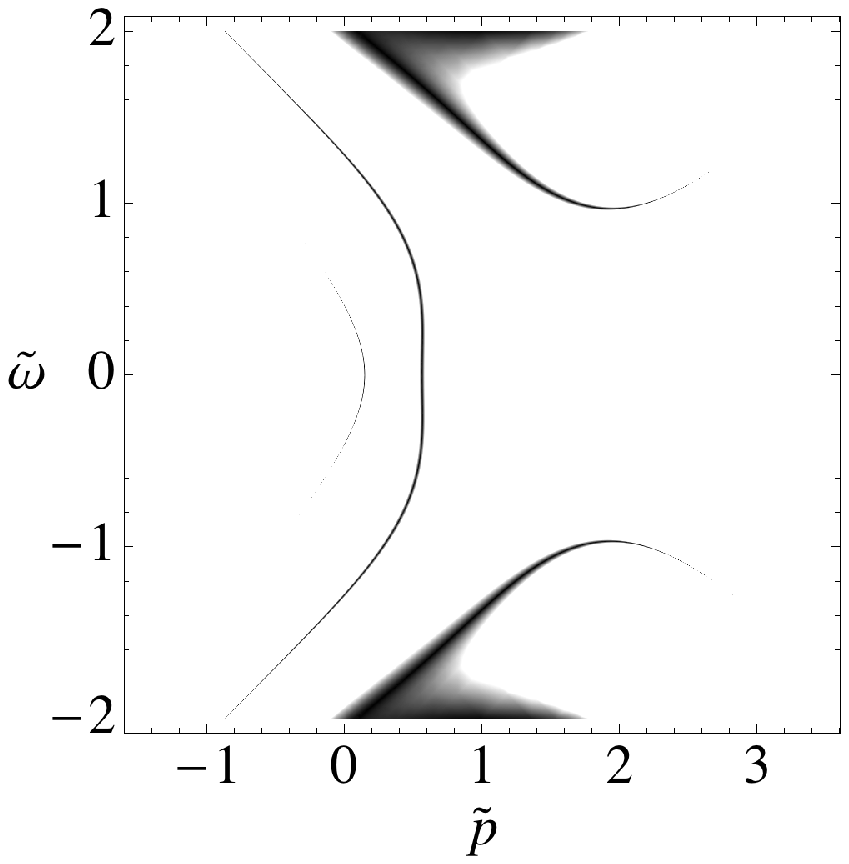}}
		\caption{\label{Fig.high} Spectral curves in $(\tp,\tw)$ at $\tB=1$ and $\tCB=0$,
		including a few higher modes too. After the lowest mode becomes unstable at $\tCA=2.03$, the next excited state becomes unstable at $\tCA=4.165$. }
\end{figure}

As a final comment,
our analysis is based on the lowest mode spectral curve instability.
There are towers of higher excited modes in the eigenvalue equation of (\ref{linear}), and in principle there will be additional higher mode instability when they become unstable.
Because these higher modes are more massive, the instability for them requires larger value of $\tCA$ than is needed for the lowest mode, and it happens always {\it within} the unstable region
already determined by
the lowest mode spectral curve.
Thus these higher mode instability
does not change the phase diagram.
See Figure \ref{Fig.high} for numerical illustrations on this point.

\section{Summary and future directions}

The fate of dense matter under extreme conditions is an interesting question relevant to high energy astrophysics and heavy-ion collisions. One such condition that we study is the application of large magnetic field, which indeed happens in the magnetars/neutron stars or off-center RHIC collisions.
Especially, the presence of strong magnetic field provides a nice arena for physics of triangle anomaly to play important dynamical roles in the system.
One of the most basic questions is the phase diagram of the favored ground state in terms of various
conditions such as temperature, baryonic/axial chemical potentials, and the strength of magnetic field, etc.
While some regions of this phase diagram are under control of perturbative QCD computations, there
are other interesting regions whose characteristic scales are much lower than $\Lambda_{QCD}$, and strong interactions are crucial in understanding their properties.
Holographic QCD based on the strong coupling expansion in large $N_c$ limit is one useful tool to
study/gain insights on some of these physics, although there are several draw-backs and limitations too.

We study the ground state of the system under a large magnetic field at zero temperature, confined/chiral-symmetry broken phase in the presence of both baryonic and axial chemical potentials.
One possible phase that was proposed by Son-Stephanov \cite{Son:2007ny} is a homogeneous distribution of axial/baryonic currents in the form of meson super-current along the direction of the magnetic field, whose magnitudes are determined exactly by triangle anomalies of QCD.
On the other hand, Basar-Dunne-Kharzeev \cite{Basar:2010zd} recently proposed an alternative ground state, at least for sufficiently large magnetic field, where
there are additional condensates of {\it transverse currents} that form spiral shapes along the magnetic field direction, and hence called {\it chiral magnetic spirals}. A basic assumption in their argument is the existence of fermionic quark quasi-particles in the effectively reduced model to 1+1 dimensions, which might be justified for sufficiently large magnetic field. However, for a more modest value of magnetic field, they are strongly coupled/confined, and it is not clear whether the picture remains effective in the strongly coupled regime. The purpose of this paper is to answer this question in the framework of holographic QCD proposed by Sakai and Sugimoto.

Our results indicate that the chiral magnetic spirals are indeed favored against the homogeneous phase at least for sufficiently large axial chemical potential. There is an issue here whether the relevant axial chemical potential that causes the phenomenon is within the validity regime of the model analysis. Especially, we used 2-derivative truncation of the DBI action on D8-branes neglecting higher derivative corrections. As our results in the previous section involve reasonable order 1 numbers in terms of dimensionless variables, we expect that our results do pass this issue safely.
However, a more thorough analysis on this aspect, particularly interesting one being whether the same conclusion will be obtained if one uses the full DBI action instead, will be a valuable future direction.

Our results do not have chiral magnetic spirals in the purely baryonic chemical potential, which in fact differs from the original proposal in Ref.\cite{Basar:2010zd} that the phenomenon should happen in both baryonic and axial cases. It would be interesting to understand why one case survives in the strong-coupling regime while the other does not.

\vskip 1cm \centerline{\large \bf Acknowledgement} \vskip 0.5cm

H.U.Y. greatly thanks Dima Kharzeev and Harmen Warringa for the invitation to the conference "Workshop on P- and CP-odd Effects in Hot and
Dense Matter", Brookhaven, April 26-30, 2010, where part of this work was presented, and also Gokce Basar and Gerald Dunne for very helpful inputs and discussions on the subject.
K.Y.K. is grateful for the support of an STFC rolling grant and would like
to thank Nick Evans, Sang-Jin Sin and Deog-ki Hong for discussions. He also would
like to thank ICTP for the hospitality during his visit, where
this work was initiated.

\appendix
\section{No chiral magnetic spirals for pure baryonic case}

In this appendix we will show rigorously that for pure baryonic case, we do not have any chiral magnetic spirals. Our starting point for this will be the master equation (\ref{linear}) replacing $F_{0Z}$ and $F_{3Z}$ with their purely baryonic expression ((\ref{background}) with $C_A=0$). Let us first expand {\it normalizable} $A^{\pm}(Z)$ in the complete basis of normalizable eigenfunctions $f_{n}(Z)$ satisfying the eigenvalue equation
\be
(1+Z^2)^{{1\over 3}}\partial_{Z}\left[(1+Z^2)\partial_Z f_{n}(Z)\right]=-\lambda_{n}^{2}f_{n}(Z)\quad,
\ee
where $0 < \lambda_{1}^{2} \leq \lambda_{2}^{2} \cdots$ and the {\it normalizable} basis functions $f_n(Z)$ satisfy the completeness condition
\be
\int (1+Z^2)^{-{1\over 3}} \bar{f_m}(Z)f_n(Z)=\delta_{nm}\quad.
\ee
These are in fact wave-functions of vector mesons one can find in Refs.\cite{Sakai:2004cn,Sakai:2005yt}.

Because $\{f_n(Z)\}$ are complete, we expand
\be
A^{\pm}(Z)=\sum_{n\geq 1} a_{n}^{\pm}f_{n}(Z)\quad,
\ee
such that $\sum_{n \geq 1}|a_{n}|^2=1$, so that $A^{\pm}(Z)$ is normalized as
\be
\int \bar{A^{\pm}}(Z) A^{\pm}(Z) (1+Z^2)^{-{1\over 3}} =1\quad.
\ee
Using the above equations and integrating the master equation in purely baryonic background after sandwiching it with $\bar{A^{\pm}}(Z)$, we get the following
\be \label{omegapb}
(\omega^2-p^2)-M_{KK}^{2}b \pm (p\alpha_1-\omega \alpha_2)=0\quad,
\ee
where $b=\sum_{n\geq 1} \left(|a_n|^2\lambda_n^{2}\right) \geq \lambda_1^{2} > 0$ and
\bear
\alpha_1 &=& {N_c C_B\over {8 \pi^2 \kappa}}\int \bar{A^{\pm}}(Z)A^{\pm}(Z){\sinh \left(\tilde{B}\tan^{-1}(Z)\right)\over{1+Z^2}}\quad, \nonumber \\
\alpha_2 &=& {N_c C_B\over {8 \pi^2 \kappa}}\int \bar{A^{\pm}}(Z)A^{\pm}(Z){\cosh \left(\tilde{B}\tan^{-1}(Z)\right)\over{1+Z^2}}\quad.
\eear
Since $\cosh \left(\tilde{B}\tan^{-1}(Z)\right) > \sinh \left(\tilde{B}\tan^{-1}(Z)\right)$, it is clear that $|\alpha_2| >|\alpha_1|$. After completing the squares in (\ref{omegapb}), we get the following
\be
\left(\omega \mp {\alpha_2 \over 2} \right)^2=\left(p\mp {\alpha_1 \over 2}\right)^2 +\left({{\alpha_2^2 -\alpha_1^2}\over 4}\right)+M_{KK}^2 b
\ee
Since all the terms in the RHS of the above equation are positive, it is clear that $\omega$ is always real in purely baryonic case and hence there is no instability and hence no signal of chiral magnetic spiral.

%%%%%%%%%%%%%%%%%%%%%%%%%%%%%%%%%%%%%%%%%%%%%%%%%%%%%%%%%%%%%%%%%%%
 \vfil

\end{document}